\documentclass{article}
\def \inbar{\vrule height1.5ex width.4pt depth0pt}
\def \C{\relax\hbox{\kern.25em$\inbar\kern-.3em{\rm C}$}}
\def \R{\relax{\rm I\kern-.18em R}}

\newcommand{\sgn}{{\rm sgn}}

\newcommand{\beq}{\begin{equation}}
\newcommand{\eeq}{\end{equation}}
\newcommand{\bea}{\begin{eqnarray}}
\newcommand{\eea}{\end{eqnarray}}
\newcommand{\nn}{\nonumber}

\begin{document}
\author{O. Teoman Turgut${\,}^{1,2,3}$\\
 \\
${}^{1}\,$Department of Physics,  Bogazici University  \\
80815, Bebek, Istanbul, Turkey\\
\\
${}^{2}\,$Department of Physics, KTH\\
SE-106 91 Stockholm, Sweden\\ 
\\
${}^{3}\,$Feza Gursey Institute\\
 81220 Kandilli, Istanbul, Turkey\\
  turgutte@boun.edu.tr\\
 turgut@theophys.kth.se}

\title{\bf On three dimensional coupled bosons }
\maketitle
\large
\begin{abstract}
\large
 In this work we study two complex   scalar fields 
coupled through a quadratic interaction  in $2+1$ dimensions using 
the method of bilinears as suggested by Rajeev\cite{2dqhd}. 
The resulting theory can be 
formulated as a classical theory. 
We study the linear approximation, and show that 
there is a possible bound  state in a range of coupling constants. 
\end{abstract}
\large
\section{Introduction}

Quantum field theory has been an essential tool for the modeling of various 
physical phenomena. 
One of the major problems in field theory is the understanding of relativistic 
bound states. The standart way to look at field theories is via perturbation
theory around the free theory, and bound state problems are difficult to 
formulate in this approach. The most common way is to use a 
Bethe-Salpeter approach for the four-point function (for two
particle bound states) and find a self-consistency
condition  for the bound state solution. Typically this requires 
various approximations which may break down in the highly relativistic
cases.
 
One of the most  successful applications of this approach is 
within the large-$N_c$ approximation: in his classic paper, 't Hooft 
obtained a bound state equation for mesons in two dimensions 
in the large-$N_c$ where $N_c$ refers to the color for the 
nonabelian $SU(N_c)$ gauge theory \cite{thooft}.
This leads to a singular integral equation for the 
possible masses of the mesonic excitations. This equation  is
expressed in terms of  
 the wave function of the meson given as a function of the fractional 
light-cone momentum.
The analysis of this integral equation  in 
\cite{hildebrant} shows that there are only bound states corresponding
to positive eigenvalues with  
finite multiplicity, and these eigenvalues  tend to infinity.
The scalar version of this model is worked out in \cite{shei} 
using the original approach of 't Hooft and in \cite{tomaras}
via a Hamiltonian approach to the large-$N_c$ limit.
These relativistic equations behave in a very similar way to the 
standart 't Hooft equation.
In two dimensions we can generalize the Yang-Mills  Lagrangian, 
since the gauge fields are not dynamical, by means of a nondynamical scalar 
field. The   large-$N_c$ limit   meson
bound state  
equation of these models have  some other interesting features \cite{douglas}.
In \cite{aoki}, Aoki has generalized these bound state equations 
for  bosons and fermions coupled  via $SU(N_c)$ gauge theory.
A good presentation of many two dimensional models using the 
bilocal fields in the path integral formalism within the 
 large-$N_c$ limit is given in \cite{cavicchi}.
In this article several interesting bound state equations are derived,
and  further references are given. 

In \cite{2dqhd} Rajeev has formulated the large-$N_c$ model
as a classical field theory  
using color invariant bilinears, and he has shown that the 
phase space of the theory is the restricted Grassmannian.
The knowledge of the phase space allows one to make a 
variational ansatz for the baryons in this theory, which correspond 
to the large fluctuations of the field(see \cite{witten}).
Further details of this approach are  given in  
\cite{istlect}.
Toprak and the author have extended this work to $SO(N_c)$ gauge 
theory of bosons and fermions and obtained  variants of the 't Hooft
equation for  these cases \cite{erdal, super}.
The adjoint matter fields in  the large-$N_c$ limit 
yields again singular  integral equations for possible 
mesonic strings, they  exhibit  a very similar  bound state 
structure as  the original model, but these equations are  
more complicated due to the fact that mesons 
are now color invariant strings of operators 
\cite{klebanov, demeterfi, kutasov, boorstein}.  
 
The two dimensional Yukawa model  is analyzed within 
the light-cone method in \cite{pauli1, pauli2, perry}.
These models are more complicated due to nonlocal renormalization effects,
it is possible to get an integral equation for   bound states using 
 some further approximations.
A four dimensional extension of these ideas are given in 
\cite{glazek}. The common feature of all these bound state equations 
is that they are singular integral equations. In the gauge theory cases these 
singular integral equations are rather restrictive in that they only 
allow for a discrete spectrum. In the other cases 
this is not necessarily 
true. There is usually a finite number (typically one) bound state.
There are investigations in three dimensional QCD for the 
bound state equations of mesons (see the recent article\cite{chakrabarti}).
Four dimensional realistic theories are very complicated since one 
has to deal with renormalization. The author is not knowledgable 
enough about these realistic bound state equations, some information 
can be found in the review article \cite{brodsky}(see also \cite{perry2}
for a review of renormalization in the light-front point of view and 
some non-perturbative applications in this formalism).

In this article we will apply a certain kind of 
mean field theory, which is  a large-$N_f$ limit 
 to  two  coupled  complex bosons 
(we call it flavor symmetry to emphasize  that it is not gauged).
This theory is simple since
it does not require coupling constant 
and wave function renormalization
in the perturbation theory.  
Defining the scalar field around the free field theory 
may not be so interesting from 
a  physical point of view, but we regard  this as an
interesting  toy model.
The linear approximation yields  a bound state for 
the composite of two bosons.
We will apply the methods of Rajeev \cite{istlect} and formulate  it as a 
classical field theory of bilinears. In this case (unlike 
the gauge theory case) this is only an 
approximation since the theory does not have to be 
restricted to this flavor invariant sector. 
To avoid repetitions we sometimes  
refer to 
our work  on complex bosons in \cite{tolyateo}.
There is an interesting  Bethe-Salpeter
treatment\footnote{I am grateful to E. Langmann for 
pointing this out to me} of bound states in the 
broken phase of $\phi^4$ theory in \cite{zarembo}, in some 
sense   this is similar to the model  we work with.

\section{The model in the light-cone and large-$N_f$ limit} 

We write down a $U(N_f)$ invariant action for two 
complex scalars $\phi_a$ and $\phi_b$, 
\beq
 S=\int d^3x \Big(\partial^\mu \phi_a^\dag\partial_\mu\phi_a
+\partial^\mu\phi_b^\dag\partial_\mu\phi_b-
m_a^2\phi^\dag_a\phi_a-m_b^2\phi^\dag_b \phi_b
-{\lambda\over 4}
[(\phi^\dag_a\phi_a)^2+(\phi_b^\dag\phi_b)^2-2(\phi_a^\dag\phi_b)
(\phi_b^\dag\phi_a)]\Big)
.\eeq
We  introduced a  common coupling $\lambda$, for  
the two complex fields $\phi_{a\alpha}$ and $\phi_{b\alpha}$.
We assume that the internal  index $\alpha$ takes the values  $1,..,N_f$.
If we do  not have the last term which couples the two fields 
the action would have a $U(N_f)\times U(N_f)$ symmetry, and 
the interaction term explicitly breaks this.
The interaction may look discomforting , but it is easy to see
that it is always positive. The classical ground state 
of the massive theory is when both fields are set to zero.
This means that we can quantize the theory around its classical minimum
by introducing creation-annihilation operators for the 
Fourier modes, as we will see below.  
Note that  
this theory in  three dimensions is super-renormalizable
from a perturbative point of view, so we do 
not expect any multiplicative renormalizations
(when $\lambda<0$ the theory 
is unstable, it can be defined by a perturbation theory analysis,
but we restrict ourselves to positive values).

Perturbatively, there is only one  type of  divergence after normal ordering
as we will comment later on \cite{peter}.
To apply the methods developed 
by Rajeev, we will use  the light-cone coordinates, 
introduce $x^+={1\over \sqrt{2}}(x^0+x^2)$ and 
$x^-={1\over \sqrt{2}}(x^0-x^2)$, and 
$x^1$ remains as the transverse coordinate.
We choose  $x^+$ as time (that is our evolution variable).
A good review of light-front methods is given in 
\cite{heinzl}, a good  discussion of the scalar field 
in the light-front is also given in \cite{yan} and in \cite{harindra}.
The three dimensional scalar field theory
has been investigated from different points of view in the 
articles  \cite{chang1, chang2, peter, windol}. 

We use basically the 
same conventions in our previous work \cite{tolyateo, rajteo}.
\bea
 S&=&\int dx^+dx^-dx^1 \Big({1\over 2}\phi_a^\dag(-2\partial_-)\partial_+\phi_a
+{1\over 2}\phi_b^\dag(-2\partial_-)\partial_+\phi_b-
\phi^\dag_a(m_a^2-\partial_1^2)\phi_a-\phi^\dag_b(m_b^2-\partial_1^2)
\phi_b\nn\cr
&\ &-{\lambda\over 4}
[(\phi^\dag_a\phi_a)^2+(\phi_b^\dag\phi_b)^2-2(\phi_a^\dag\phi_b)
(\phi_b^\dag\phi_a)]\Big)
.\eea
The action is first order in time $x^+$, this means that we are
already in the Hamiltonian formalism. 
The Hamiltonian can be read off directly,
\beq
    H=\int dx^-dx^1\Big(
\phi_a^\dag(-\partial_1^2+{m_a^2})\phi_a
+\phi^\dag_b(-\partial_1^2+m_b^2)\phi_b+{\lambda\over 4}
[(\phi^\dag_a\phi_a)^2+(\phi_b^\dag\phi_b)^2-2(\phi_a^\dag\phi_b)
(\phi_b^\dag\phi_a)]\Big)
.\eeq

The quantization at equal time following Dirac gives 
\beq
    [\hat \phi_a^{\alpha\dag}(x^-,x^1), \hat \phi_{a\beta}(y^-, y^1)]=
-{i\over 4} \delta^{\alpha}_{\beta}
\sgn(x^--y^-)\delta(x^1-y^1)
,\eeq
and the same rule applies for $\phi_b$.
We recall that the field can be expanded in 
terms of creation-annihilation operators 
at initial light-front time (since the classical minimum is 
the zero configuration for the fields),
\bea
\hat \phi_{a\alpha}(x^-,x^1)&=&\int {[dp_-dp_1]\over \sqrt{2|p_-|}}
a_\alpha(p_-,p_1)e^{-ip_-x^--p_1x^1}\nn\cr
 \hat \phi_{b\alpha}(x^-,x^1)&=&\int {[dp_-dp_1]\over \sqrt{2|p_-|}}
b_\alpha(p_-,p_1)e^{-ip_-x^--p_1x^1} 
,\eea  
where we use $[dp]={dp\over 2\pi}$.
(To properly define everything we should assume that 
this expansion is given for $(-\infty,-\epsilon_0]\cup[\epsilon_0,\infty)$
at the end we take $\epsilon_0\to 0$ limit).

The creation and annihilation operators now satisfy,
\bea
    [a_\alpha(p_-,p_1),a^{\beta\dag}(q_-,q_1)]&=&
\sgn(p_-)\delta^\beta_\alpha\delta[p_- -q_-]\delta[p_1-q_1]
,\nn\cr
 [\,b_\alpha(p_-,p_1),b^{\beta\dag}(q_-,q_1)]&=&
\sgn(p_-)\delta^\beta_\alpha\delta[p_- -q_-]\delta[p_1-q_1]
.\eea
Since the fields are complex valued 
 annihilation and creation operators are not related.
We introduce a  vacuum state $|0>$ for the 
Fock space construction,
\bea
   a_\alpha(p_-,p_1)|0>=0\  \ {\rm for}\  p_->0 \ \ {\rm and}\  \ 
 a^{\alpha\dag}(p_-,p_1)|0>=0 \  \ {\rm for} \  p_-<0\nn\cr 
b_\alpha(p_-,p_1)|0>=0\  \ {\rm for}\  p_->0 \ \ {\rm and}\  \ 
 b^{\alpha\dag}(p_-,p_1)|0>=0 \  \ {\rm for} \  p_-<0
.\eea
It is important to keep in mind that 
the operator $a_\alpha(p_-,p_1)$ for $p_-<0$, {\it creates an antiparticle
of momentum } $(-p_-, -p_1)$ and similarly for $b_\alpha$
(which one can see by rewriting the above expansions in a more
conventional  way, by separating particle and antiparticle 
operators).
We define  normal ordering rules with respect to this 
vacuum  as usual and denote it by 
a colon $:\phi_a...\phi_a:$, we will be using  
for the  computations the following relation,
\beq
  :a^{\alpha\dag}(p_-,p_1)a_\beta(q_-,q_1):=
a^{\alpha\dag}(p_-,p_1)a_\beta(q_-,q_1)
-{\delta^\alpha_\beta\over 2}(1-\sgn(p_-))\delta[p_--q_-]\delta[p_1-q_1]
,\eeq
and exactly the same for $b$ quanta.

We have the following Hamiltonian in the 
quantized theory, 
\bea
   \hat H&=&\int dx^1dx^-\Big( 
: \hat \phi_a^{\dag}(m_a^2 -\partial_1^2)\hat\phi_{a}:
+: \hat \phi_b^{\dag}
(m_b^2 -\partial_1^2)\hat\phi_b:\nn\cr
&+&{\lambda\over 4}
[:(\hat \phi^\dag_a\hat \phi_a)^2:+:(\hat \phi_b^\dag\hat \phi_b)^2:
-2:(\hat \phi_a^\dag\hat \phi_b)
(\hat \phi_b^\dag\hat \phi_a):]\Big)
\eea
As it stands the Hamiltonian would not be a well-defined operator
for finite $N_f$ theory, we need to introduce  mass renormalization  
terms  which correspond in the diagramatic language 
the setting-sun diagrams\cite{chang1, chang2, peter}.
When  we take the large-$N_f$ limit  these 
counter terms become of smaller order, therefore the 
Hamiltonian as written will have a well-defined limit.

We now define as an approximation a  large-$N_f$ limit
and restrict the theory to the flavor invariant sector. 
This is to be taken as an approximation to the full quantum theory. 
We  introduce a set of 
flavor invariant operators, which are directly written in the
momentum representation, and to
simplify  notation we
write $p$ to denote $p_-, p_1$ collectively, and 
$\delta[p-q]=\delta[p_--q_-]\delta[p_1-q_1]$,
\bea
\hat N_a(p,q)&=&{2\over N_f}:a^{\alpha\dag}(p_-,p_1)a_\alpha(q_-,q_1):\,\cr
\hat N_b(p,q)&=&{2\over N_f}:b^{\alpha\dag}(p_-,p_1)b_\alpha(q_-,q_1):\,\cr
\hat C(p,q)&=& {2\over N_f}\,a^{\alpha\dag}(p_-,p_1)b_\alpha(q_-,q_1)\,\cr
\hat {\bar C}(p,q)&=&{2\over N_f}\,b^{\alpha\dag}(p_-,p_1)a_\alpha(q_-,q_1).
\eea
Note that  $\hat C$ and $\hat {\bar C}$ are just hermitian 
conjugates of each other.

The idea behind the papers \cite{2dqhd, istlect} is that when 
we take the large-$N_f$ limit the flavor invariant 
operators have smaller and smaller fluctuations, and if we compute
their commutator, for example for $N_a$ with itself,
we get,
\bea
[\hat N_a(p,q), \hat N_a(s,t)]&=&{2\over N_f}\Big(\hat N_a(p,s)\sgn(p_-)
\delta[q-r]-\hat N_a(r,q)\sgn(p_-)\delta[p-s]\nn\cr
 &-&(\sgn(p_-)-\sgn(q_-))\delta[p-s]\delta[q-r]\Big)
.\eea
We assume that when we let $N_f\to \infty$ there are proper large-$N_f$ limits 
for these bilinears restricted to the flavor invariant states.
As a result the theory becomes classical, the expectation values 
of flavor invariant operators factorize as $N_f\to \infty$ 
\cite{istlect, yaffe}.
Thus we may postulate a set of Poisson brackets for these 
{\it classical} variables:
\bea
   \{N_a(p,q), N_a(s,t)\}&=&-2i\Big(N_a(p,s)\sgn(p_-)\delta[q-r]-
N_a(r,q)\sgn(p_-)\delta[p-s]\nn\cr
&\ & -(\sgn(p_-)-\sgn(q_-))\delta[p-s]\delta[q-r]\Big)\cr
\{N_b(p,q), N_b(s,t)\}&=&-2i\Big(N_b(p,s)\sgn(p_-)\delta[q-r]-
N_b(r,q)\sgn(p_-)\delta[p-s]\nn\cr
&\ & -(\sgn(p_-)-\sgn(q_-))\delta[p-s]\delta[q-r]\Big)\cr
 \{C(p,q), \bar C(s,t)\}&=&-2i\Big(N_a(p,t)\sgn(q_-)\delta[q-s]
-N_b(s,q)\sgn(t_-)\delta[p-t]\cr
&\ &-(\sgn(p_-)-\sgn(q_-))\delta[p-t]\delta[q-s]\cr
\{N_a(p,q), C(s,t)\}&=&-2i \sgn(q_-)\delta[q-s]C(p,t)\cr
\{N_b(p,q), C(s,t)\}&=&2i\sgn(q_-)\delta[p-t]C(s,q)\cr
\{N_a(p,q), \bar C(s,t)\}&=&2i\sgn(p_-)\delta[p-t]\bar C(s,q)\cr
\{N_b(p,q), \bar C(s,t)\}&=&-2i\sgn(q_-)\delta[q-s]\bar C(p,t)
,\eea
and all the other Poisson brackets being zero.

There are constraints coming from the $U(N_f)$ invariance, if 
we restrict the theory to the flavor invariant sector.
They are very similar to the ones found in \cite{tolyateo},
\bea
&\ &(\epsilon N_a+\epsilon)^2+\epsilon C\epsilon C^\dag=1,\nn\cr
&\ & (\epsilon N_b+\epsilon)^2+\epsilon C^\dag \epsilon C=1,\nn\cr
&\ & N_a\epsilon C+C\epsilon N_b+\epsilon C+C\epsilon =0,
\eea
and the hermitian conjugate of the last 
equation. These conditions  
are derived using similar techniques to \cite{rajteo,tolyateo}.
Here we are 
using 
$\epsilon(p,q)=\sgn(p_-)\delta[p_--q_-]\delta[p_1-q_1]$, 
$1(p,q)=\delta[p-q]$,
and a shorthand for the 
operator products, for example, $N_aC$ means,
\beq
   (N_aC)(p,s)=\int [dq_-dq_1]N_a(p_-,p_1;q_-,q_1)C(q_-,q_1;s_-,s_1)
,\eeq
and similarly for the others.

We will have further convergence conditions coming from the 
super-renormalizability. These should be  regarded as sufficiently restrictive
conditions to keep the system's evolution in phase space. The Hamiltonian 
puts more stringent conditions on the admissible class of observables, we 
believe  
its domain should be dense inside the phase space.
Correct normalization should be then found using the Hamiltonian as 
a quadratic form on the space of these variables and demanding this
form 
to be finite for all physical states. 
To state our conditions we 
recall that the one-particle Hilbert spaces of bosons  are
divided into positive and negative energy subspaces
according to $\sgn(p_-)$.
We assume that the operators which act between 
positive and negative energy subspaces will be 
Hilbert-Schmidt, whereas the operators acting between the 
same subspaces will be trace class \cite{berezin}. 
We write explicitly the one for $C$:
$C(u_-,u_1;v_-,v_1)$ is trace class for 
$u_-v_->0$, and Hilbert-Schmidt for $u_-v_-<0$, and the 
same for the other variables. Note that these are
consistent with the constraints on the system, which 
defines the geometry of the phase space.
The constraints and the convergence conditions can be 
cast into a coherent geometric picture: it defines a
a homogeneous manifold of 
$U_1(({\cal H}_a \oplus {\cal H}_b)_+,({\cal H}_a \oplus {\cal H}_b)_-)$, but 
we will not use it in this work.

We can rewrite the large-$N_f$ Hamiltonian
in terms of these variables,
\bea
   H&=&{1\over 4}\int [dp]{m_a^2+p_1^2\over |p_-|}N_a(p,p)+
{1\over 4}\int [dp]{m_b^2+p_1^2\over |p_-|}N_b(p,p)\nn\cr
&+& {\lambda\over 64}\int [dpdqdsdt]{\delta[p-q+s-t]\over
\sqrt{|p_-q_-s_-t_-|}}
\Big[N_a(p,q)N_a(s,t)+N_b(p,q)N_b(s,t)-2C(p,q)\bar C(s,t)\Big]
\eea

This defines our large-$N_f$ approximation, and in principle 
we can calculate the equations of motion of the basic 
observables by,
\beq
   {\partial O(u,v)\over \partial x^+}=\{ O(u,v), H\}
,\eeq
where $O$ refers to any one of $N_a, N_b, C, \bar C$.
The resulting equations are  nonlinear integral equations 
and we also have the constraint to satisfy.
It would be interesting to study this system using a variational 
ansatz. We will leave the analysis of the full system to a future work
and look at a linearized version.

\section{Linearization and a possible bound   state}

To get a better feeling about the system we can start with a 
linear approximation. 
This means we should linearize the constraint as well as the equations 
of motion.
The linearization of the constraint gives us, for our basic 
variables,
 \bea
   (1+\sgn(u_-)\sgn(v_-))N_a(u_-,u_1;v_-,v_1)&=&0\quad
(1+\sgn(u_-)\sgn(v_-))N_b(u_-,u_1;v_-,v_1)=0\cr
(\sgn(u_-)+\sgn(v_-))C(u_-,u_1,v_-,v_1)&=&0
.\eea
We will be using  the last one in our computations,
it says that the light-cone momenta should be 
opposite to each other:
$C(u_-,u_1;v_-,v_1)=0$  if $u_-v_->0$, and nonzero
if $u_-v_-<0$. The same conditions hold for 
$N_a$ and $N_b$ as well.
In the linear approximation we will search for 
a possible bound state 
of $a$ and $b$ particles. 
In principle we can compute the linearized equations
for all the other variables, but they will not lead to a solution 
for the bound state nor a solution for a resonance: they 
only have scattering states. 
Therefore we work only with the 
composite $C(u,v)$, 
let us choose
$u_->0,v_-<0$. 
The equations of motion of $C(u,v)$ for $u_->0, v_-<0$, 
in the linear approximation become,
\bea
   {\partial C(u,v)\over \partial x^+}&=&
\{C(u,v), H\}={i\over 2}\Big[{m_a^2+u_1^2\over u_-}-{m_b^2+v_1^2\over
v_-}\Big]C(u,v)\nn\cr
&-&i{\lambda\over 8\pi}\int {[dpdq]\over \sqrt{|p_-q_-u_-v_-|}}
\delta[p_--q_-+v_--u_-]\delta[p_1-q_1+v_1-u_1]
C(u,v)
\eea

Let us make an ansatz,
in the light-front  direction we make a `t Hooft like
choice with respect to the relative momentum variable 
$\zeta=u_-/(u_--v_-)$. This variable now satisfies $0~<~\zeta~<~1$, and we set 
$C(u_-,u_1;v_-,v_1) =\tilde f(\zeta;u_1, v_1)e^{iP_+x^+}$.
Notice that $P_-=u_--v_->0$, and
we introduce  
 a relativistically  invariant mass variable 
$\mu^2=2P_+|u_--v_-|-(u_1+(-v_1))^2$, 
which will be the mass of the bound state
(recall that the momentum $v_1$ denotes an antiparticle with
momentum $-v_1$, thus $u_1+(-v_1)$ is the 
total transversal momentum of the 
two particle state). 
After some manipulations, similar to the ones in \cite{2dqhd, erdal}, 
this gives us an eigenvalue equation for the invariant mass: 
\bea
  \mu^2\tilde f(\zeta;u_1,v_1)&=&
\Big[{m_a^2 +(u_1-\zeta (u_1+(-v_1)))^2\over \zeta}+
{m_b^2+(v_1-(1-\zeta)(u_1+(-v_1)))^2
\over 1-\zeta}\Big]\tilde f(\zeta;u_1,v_1)\nn\cr
&-&{\lambda \over 8\pi}\int_0^1 d\eta \int_{-\infty}^{\infty} 
{[dp_1dq_1] \tilde f(\eta; p_1,q_1)\over \sqrt{\eta(1-\eta)\zeta(1-\zeta)}}
\delta[p_1-q_1-(u_1+(-v_1))]
.\eea
(Note that this form reduces to $\mu=m_a+m_b$ if we set $\lambda=0$, and 
choose the function $\tilde f$ properly).
We may equivalently use a new set of variables,
$R=u_1-\zeta (u_1+(-v_1))$ and $u_1+(-v_1)$, 
relative transversal light-front momentum and transversal total momentum, 
respectively,  instead of the above 
variables. If we write everything in terms of these 
new set of variables we get,
\bea
  \mu^2\tilde f(\!\!\!\!&\zeta&\!\!\!;R, u_1+(-v_1))=
\Big[{m_a^2 +R^2\over \zeta}+{m_b^2+(-R)^2
\over 1-\zeta}\Big]\tilde f(\zeta;R, u_1+(-v_1))\nn\cr
&-&{\lambda \over 8\pi}\int_0^1 d\eta \int_{-\infty}^{\infty} 
{[dQ][d(p_1-q_1)] \tilde f(\eta; Q,p_1+(-q_1))\over 
\sqrt{\eta(1-\eta)\zeta(1-\zeta)}}
\delta[p_1-q_1-(u_1+(-v_1))]
.\eea
(The Jacobian of this transformation in the integral is one), 
The total momentum integral can be done due to the delta 
function and we end up with,
\bea
  \mu^2\tilde f(\zeta;R, u_1+(-v_1))&=&
\Big[{m_a^2 +R^2\over \zeta}+{m_b^2+(-R)^2
\over 1-\zeta}\Big]\tilde f(\zeta;R, u_1+(-v_1))\nn\cr
&-&{\lambda \over 8\pi}\int_0^1 d\eta \int_{-\infty}^{\infty} 
{[dQ]\tilde f(\eta; Q,u_1+(-v_1))\over 
\sqrt{\eta(1-\eta)\zeta(1-\zeta)}}
.\eea
We see that the total momentum 
$u_1+(-v_1)$ is conserved, thus  we can factor out  
the transversal center of mass motion 
by assuming $\tilde f(\zeta;u_1, v_1)=f(\zeta, R)g(u_1+(-v_1))$:
\beq
  \mu^2 f(\zeta, R)=\Big[ {m_a^2+R^2\over \zeta}+
{m_b^2+R^2\over 1-\zeta}\Big]f(\zeta, R)
- {\lambda\over 16\pi^2}\int_{-\infty}^{\infty}dQ\int_0^1 d\eta {f(\eta, Q)
\over \sqrt{\eta(1-\eta)\zeta(1-\zeta)}}
.\eeq  
We can reduce this again to a functional equation for the 
unknown eigenvalue, by using the standart techniques,
\beq
   {\lambda\over 16\pi^2}\int_{-\infty}^{\infty}\int_0^1 {d\eta
dQ \over R^2+m_a^2+\mu^2\eta^2+(m_b^2-m_a^2-\mu^2)\eta}=1
.\eeq
The integrand will have no poles if the quadratic expression involving
$\eta$ has no real roots, or if it has a double root.
This is the case if 
$|m_a-m_b|\leq \mu\leq m_a+m_b$.
We assume this case, the last one is clear it says that the 
bound state cannot be of bigger mass, the 
other one says that the fundamental quanta should be stable 
against decay, if for example $m_a>m_b$
then it would be favorable to have
$m_a\mapsto m_b+\mu$.
Then we can evaluate the integral in any way we want, first we take the 
$Q$ integral, this gives us,
\beq
    \int_0^1 {d\eta\over \sqrt{\mu^2\eta^2+(m_b^2-m_a^2-\mu^2)\eta
+m_a^2}}={16\pi\over \lambda}
.\eeq
The next integral can be done and simplified into 
\beq
   {1\over \mu}\ln\Big[{m_a+m_b+\mu\over m_a+m_b-\mu}\Big]=
{16\pi\over \lambda}
,\eeq
which is valid when $|m_a-m_b|\leq \mu \leq m_a+m_b$.
We may study the 
small coupling limit of this expression.
In this case 
we expect that  the bound state mass  becomes very close to the two mass
treshold, then we can 
write 
\beq 
{m_a+m_b-\mu \over m_a+m_b+\mu}\approx e^{-{16\pi (m_a+m_b)/ \lambda}},
\quad \mu \approx (m_a+m_b)(1-2e^{
-{16\pi (m_a+m_b)/ \lambda}})
,\eeq
which is 
consistent if we take 
$\lambda/(m_a+m_b)<<1$.
the other extreme is interesting as well, 
$\mu\approx m_a-m_b$ (assuming $a$ is the heavier particle),
this implies a critical coupling $\lambda_c$ 
 beyond which our methods break down, due to 
the appearance of a tachyon,
\beq
{16\pi\over \lambda_c}={1\over m_a-m_b}\ln\Big[{m_a\over m_b}\Big]
\quad {\rm or} \quad \lambda_c={16\pi\over \ln[{m_a\over m_b}]}(m_a-m_b)
.\eeq
This critical 
coupling is pushed to higher and higher values if 
$b$ becomes lighter and lighter 
with respect to 
the $a$ particle. 
For any given value of the 
coupling constant in the interval $(0, \lambda_c]$,
there is a solution for the bound state energy.
Hence we see that there is a composite 
bound state for these values of the coupling constants.
It is not clear what happens beyond this value, it is possible that 
the linear approximation breaks down, it is also possible that the
large-$N_f$ limit is not a good approximation beyond a 
certain value. The other possibility is that the naive vacuum is 
not a true vacuum of the 
quantum theory and we should redefine the vacuum of the system.
We are not able to analyze these possiblities at the moment.

Let us  compare this with the results we would have found if we 
looked at a $1+1$-dimensional version of the same model.
Then the bound state equation could be written in terms of 
the fractional light-cone momentum $\zeta$ only.
There is no transversal component and we have only one 
integral to compute.
As a result we find the equation that should be 
satisfied by the eigenvalue $\mu$, (with the
condition $|m_a-m_b|\leq \mu\leq m_a+m_b$),
\beq
{1\over \sqrt{4m_a^2m_b^2-z^2}}\Big[\arctan\Big({2m_b^2+z\over 
\sqrt{4m_a^2m_b^2-z^2}}\Big)+\arctan\Big({2m_a^2+z\over 
\sqrt{4m_a^2m_b^2-z^2}}\Big)\Big]={8\pi\over \lambda^2}
,\eeq
where $\mu^2=m_a^2+m_b^2+z$ and we wrote $\lambda^2$ for the 
coupling constant since it has dimensions mass-squared.
To analyze the behaviour it is more natural to define 
the dimensionless variables,
$\tilde z=z/2m_am_b$, and   $\sigma={m_a/m_b}$, and rescale 
the coupling $\tilde \lambda^2=\lambda^2/m_am_b$,
\beq
{1\over \sqrt{1-\tilde z^2}}\Big[\arctan\Big({\sigma+\tilde z\over 
\sqrt{1-\tilde z^2}}\Big)+\arctan\Big({1/\sigma+\tilde z\over 
\sqrt{1-\tilde z^2}}\Big)\Big]={16\pi\over \tilde \lambda^2}
.\eeq
Note that now $\tilde z$ satisfies $-1<\tilde z<1$.
If we take the limit $\tilde z\to -1^+$, this 
corresponds to $\mu\to |m_a-m_b|^+$ and 
the other limit $\tilde z\to 1^-$ corresponds 
to $\mu\to (m_a+m_b)^-$.
If we assume $\tilde z\approx 1^-$ we see that the 
bound states satisfy a relation
\beq 
   \tilde z\approx 1-{\tilde \lambda^4\over 128}, \quad {\rm or}\quad
\mu\approx (m_a+m_b)\Big[1-{\lambda^4\over 128 m_am_b(m_a+m_b)^2}\Big]
.\eeq
and if we take $\tilde \lambda$ sufficiently small this is 
consistent.
Notice that in $2+1$ dimensions we have an exponential 
behaviour in the inverse coupling, which is nonanalytic in the
coupling constant (around zero), as opposed to this power law change.
If we look at the other extreme we see that 
there is a finite limit for $\tilde z\to -1^+$, in fact 
it is equal to $1$. This implies a critical coupling 
again, beyond which our methods predict a tachyonic state,
\beq
\tilde \lambda_c^2=16\pi, \quad {\rm or} \quad \lambda^2_c=16\pi m_am_b.
.\eeq
This is to be compared  with the result in equation (14), 
which is sensitive to the mass difference.

Let us  comment on the convergence conditions in this 
context.
Since we are looking for a normalizable  solution it looks natural to  demand 
\beq
     \int_0^1 d\zeta \int_{-\infty}^\infty  [dR] |f(\zeta, R)|^2<\infty
.\eeq   
In fact this is  right, and we could see this from 
 our Hilbert-Schmidt condition,
\beq
    \int_{u_-v_-<0} [du_-dv_-][du_1dv_1]|C(u_-,u_1;v_-,v_1)|^2<\infty
.\eeq
If we now make the above change of variables by calling $u_--v_-=P_-$
we have
\beq
    {1\over \pi}\int_0^1 d\zeta
\int_{0^+}^\infty P_-[dP_-]\int [d(u_1+(-v_1))dR]
|C(P_-,\zeta;R, (u_1+(-v_1)))|^2<\infty,
\eeq
In our case we are  restricting $P_-$ to  the surface 
$2P_-P_+=\mu^2+(u_1+(-v_1))^2$ 
(for fixed $\mu, P_+$), this means we should 
reinterpret   the 
above normalization as  
\beq
    \int_0^1  d\zeta \int [dR] |f(\zeta,R)|^2 \int {\mu^2+(u_1+(-v_1))^2\over
2P_+}|g(u_1+(-v_1))|^2 [d(u_1+(-v_1))]<\infty
,\eeq
(notice that $P_+$ is not allowed to be zero), which means two 
separate conditions,
\beq
\int_0^1  d\zeta \int [dR] |f(\zeta,R)|^2<\infty\quad 
 \int {\mu^2+(u_1+(-v_1))^2\over
2P_+}|g(u_1+(-v_1))|^2 [d(u_1+(-v_1))]<\infty
.\eeq
The second  one  simply is a Sobolev type 
condition, which states  that the  energy of the 
transversal center of momentum  component should also be finite.
We see that our solution actually satisfies  a stronger 
condition for equations of motion to make sense.

\section{Acknowledgements} The author would like to thank Rajeev
for several ideas related to the scalar field theory and large-$N_f$ 
limits. The anonymous referee  pointed out 
several errors in the first version which dealt with the
ordinary scalar self-coupled  field, the present 
version has been written following the referee's 
careful critisms and suggestions.
The author   gratefully acknowledges the referee's 
suggestions and  discussions with M. Arik, E. Langmann, 
J. Mickelsson, C. Saclioglu, and 
E. Toprak.   The author's stay in KTH is made possible by the 
Gustafsson Fellowship, and he would like to thank to J. Mickelsson for this
great opportunity.


\begin{thebibliography}{35}

\bibitem{thooft}  G. `t Hooft, {\it A two dimensional model for 
mesons}, Nuc. Phys.  {\bf B 75} (1974) 461.

\bibitem{hildebrant} S. Hildebrandt and V. Visnjic,
{\it Meson wave functions in two-dimensional quantum chromodynamics},
Phys. Rev. {\bf D 17} (1978) 1618.

\bibitem{shei} S. S. Shei  and H-S. Tsao, {\it Scalar quantum Chromodynamics 
in two dimensions and the parton model}, Nucl. Phys. {\bf B 141} (1978) 445.

\bibitem{tomaras} T. N. Tomaras, {\it Scalar U(N) QCD in the 
large-N limit}, Nuc.  Phys.{\bf  B 163} (1980) 79.

\bibitem{douglas} M. R. Douglas, K. Li, and M. Staudacher,
{\it Generalized two dimensional QCD}, Nucl. Phys. {\bf 420} (1994)
118.


\bibitem{aoki}  K. Aoki, {\it Boson-fermion bound states in two dimensional 
QCD}, Phys. Rev. {\bf D 49} (1994) 573,
 K. Aoki and T. Ichihara, {\it 1+1 dimensional QCD with fundamental bosons and fermions}
Phys. Rev. {\bf D 52} (1995) 6453.

\bibitem{cavicchi}  M. Cavicchi, {\it A bilocal field approach to
the large-N expansion of two dimensional (gauge) theories}, Intr. Jour. Mod. Phys. {\bf A 10} 
(1995) 167.


\bibitem{2dqhd} S. G. Rajeev, {\it Quantum hadrondynamics in two dimensions},
Int. J. Mod. Phys.{\bf A 9} (1994) 5583.

\bibitem{witten}E. Witten, {\it Baryons in the 
1/N expansion}, Nucl. Phys. {\bf B 160} (1979) 57.

 \bibitem{istlect} S. G. Rajeev, {\it Derivation of  Hadronic Structure  Functions from QCD},
Conformal Field Theory, edited by Y. Nutku, C. Saclioglu, O. T. Turgut,
Perseus Publishing, New York, 2000.

\bibitem{erdal} E. Toprak and O. T. Turgut, {\it large N limit of
SO(N) scalar gauge theory}, Jour. Math. Phys   {\bf 43 }   (2002) 1340.

\bibitem{super} E. Toprak and O. T. Turgut, {\it large N limit 
of SO(N) coupled bosons and fermion}, to be published in JMP.



\bibitem{klebanov} S. Dalley and I. Klebanov,
{\it String Spectrum of 1+1-Dimensional Large N QCD with Adjoint Matter},
 Phys. Rev. {\bf D 47} (1993) 2517. 

\bibitem{demeterfi} G. Bhanot, K. Demeterfi,  and I. Klebanov, 
{\it $1+1$-Dimensional Large $N$ QCD coupled to Adjoint Fermions}, 
Phys. Rev. {\bf D 48} (1993) 4980.

\bibitem{kutasov} D. Kutasov, {\it Two Dimensional QCD coupled to 
Adjoint Matter and String Theory},
 Nucl. Phys. {\bf B 414} (1994) 33.

\bibitem{boorstein} J. Boorstein and D. Kutasov,
{\it Symmetries and Mass Splittings in QCD$_2$ Coupled to Adjoint Fermions}, 
Nucl. Phys. {\bf B 421} (1994) 263.

\bibitem{pauli1} H-C. Pauli and S. J. Brodsky, {\it 
Solving field theory in one space and one time dimension},
Phys. Rev. {\bf D 32},
(1985) 1993.
 
\bibitem{pauli2} H-C. Pauli and S. J. Brodsky, {\it Disceretized light-cone 
quantization: solution to a field theory in one space and one time 
dimension}, Phys. Rev. {\bf D 32}, (1985)
2001.

\bibitem{perry} R. J. Perry and A. Harindranath,
{\it Renormalization in the light-front Tamm-Dancoff approach to field 
theory}, Phys. Rev. {\bf D 43}, (1991)
4051.

\bibitem{glazek} S. Glazek, A. Haindranath, S. Pinsky, J. Shigemitsu, 
and K. Wilson, {\it Relativistic bound-state probelm in the light-front 
Yukawa model}, Phys. Rev. {\bf D 47} (1993) 1599.




\bibitem{chakrabarti} D. Chakrabarti and A. Harindranath, {\it 
Mesons in the light-front $QCD_{2+1}$: investigation of Bloch effective 
Hamiltonian}, Phys. Rev. {\bf D 64} (2001) 105002.

\bibitem{brodsky} S. Brodsky, H-C. Pauli and S. Pinsky, {\it QCD 
and other field 
theories on the light cone} Phys. Report. {\bf 301} (1998) 299. 

\bibitem{perry2} R. J. Perry,
{\it A renormalization group approach to Hamiltonian light-front field theory}, Annals Phys. 232 (1994) 116
 

\bibitem{tolyateo} A. Konechny and O. T. Turgut, {\it Supergrassmannian 
and large N limit of quantum field theory with bosons and fermions},
to be published in JMP.
 
\bibitem{zarembo}M. Caselle, M. Hasenbusch, P. Provero and K. Zarembo,
{\it Bound states and glue balls in three-dimensional Ising systems},
{\bf hep-th/0103130}.

\bibitem{peter} A. Peter, J. M. Haueser, M. H. Thoma, and 
W. Cassing, {\it Cluster expansion approach to the effective potential
in $\phi^4_{2+1}$ theory}, Z. Phys. {\bf C 71} (1996) 515.


\bibitem{heinzl} T. Heinzl, {\it Light cone quantization: foundations and applications},
in {\it methods of quantization}, Proceedings of 39th Schlading winter school.
{\bf hep-th 0008096}.
 
\bibitem{yan} S-J. Chang, R. G. Root, and T-M. Yan, 
{\it Quantum field theories in infinte momentum frame. I. Quantization of 
scalar and Dirac fields}, Phys. Rev. {\bf D 7} (1973) 1133.  

\bibitem{harindra} A. Harindranath, {\it An introduction to light front
 dynamics for pedestrians },
Published in Light-Front Quantization and
      Non-Perturbative QCD, J.P. Vary and F. Woelz (eds.), 
International Institute of Theoretical
      and Applied Physics, ISU, Ames, IA 50011, U.S.A. ISBN: 1-891815-00-8

\bibitem{chang1} S-J. Chang and J. A. Wright, {\it Quantum fluctuations 
in a $\phi^4$  field theory. II. One mode approximation}
Phys. Rev. {\bf D 12} (1975) 1595.

\bibitem{chang2} S-J. Chang and S. F.  Magruder,{\it Normal operators 
and Hamiltonians in $(\phi^4)_3$ field theory},
Phys. rev. {\bf D 16} (1977) 983.


\bibitem{windol} M. Windolowski, {\it A Nonperturbative study of 
three dimensional $\phi^4$ theory}, {\bf hep-th 0002243}. 

\bibitem{rajteo} S. G. Rajeev and O. T. Turgut, {\it 
Geometric quantization and two dimensional QCD},
Comm. Math. Phys. {\bf 192} (1998) 493.
 
\bibitem{yaffe} L. Yaffe, {\it Large-N limits as classical mechanics}, 
Rev. Mod. Phys. {\bf 54} (1982) 407.


\bibitem{berezin} A. F. Berezin, {\it The method of second quantization},
Academic Press (1968).

\end{thebibliography}
\end{document}